\documentclass[aps,prl,preprint,superscriptaddress,floatfix]{revtex4}

\usepackage{graphicx,latexsym}

\begin{document}


\title{Transition from Icosahedral to Decahedral Structure in a Coexisting Solid-Liquid Nickel Cluster}


\author{D. Schebarchov}
\affiliation{MacDiarmid Institute for Advanced Materials
and Nanotechnology, School of Chemical and Physical Sciences,
Victoria University of Wellington, New Zealand}
\author{S. C. Hendy}
\affiliation{MacDiarmid Institute for Advanced Materials
and Nanotechnology, School of Chemical and Physical Sciences,
Victoria University of Wellington, New Zealand}
\affiliation{Industrial Research Ltd, Lower Hutt, New Zealand}


\date{\today}

\begin{abstract}
We have used molecular dynamics simulations to construct a microcanonical caloric curve for a 1415-atom Ni icosahedron. Prior to melting the Ni cluster exhibits static solid-liquid phase coexistence. Initially a partial icosahedral structure coexists with a partially wetting melt. However at energies very close to the melting point the icosahedral structure is replaced by a truncated decahedral structure which is almost fully wet by the melt. This structure remains until the cluster fully melts. The transition appears to be driven by a preference for the melt to wet the decahedral
structure. 
\end{abstract}


\maketitle


It is well known that in small fcc clusters, noncrystalline structures such as icosahedra and decahedra can become stable as the total surface energy becomes comparable to the energy of the interior \cite{Ino66}. Similarly, in small coexisting solid-liquid clusters \cite{Pochon04, Lee04} the energy of the solid-liquid interface makes a substantial contribution to the total energy of the cluster, and can strongly influence cluster properties \cite{Honeycutt87, Reiss88, Wales94}. For example, to avoid the cost of forming the interface, small clusters tend to avoid static coexistence. This leads to an S-bend in the microcanonical caloric curve \cite{Lynden-Bell}, and the corresponding negative heat capacities observed in small sodium clusters \cite{Schmidt01}. In fact, in sufficiently small clusters static solid-liquid coexistence may not occur at all \cite{Hendy05} allowing only a dynamic coexistence between the solid and liquid phase \cite{Honeycutt87}. However, in larger clusters, just above this threshold, where the cost of forming the interface is significant but not prohibitive, it may be that the solid component assumes new structures in order to form a more favorable interface with the liquid component. This would have interesting consequences for the interpretation of premelting features in cluster caloric curves \cite{Breaux05}, and may also offer new ways of controlling cluster structures. 

Indeed, the delicate balance of internal and surface energies that produce icosahedra and decahedra is further upset in 
a coexisting cluster by the fact that the solid region is reduced in size. While the fully solid cluster
may be in the size range where one structure is stable, the smaller solid cluster coexisting with the liquid may
be in a size range where another structure is stable. It is possible that a rich variety of structures may be observed in coexisting solid-liquid clusters as new energetic balances are established, and it is not difficult to imagine a sequence of structural transitions as the liquid fraction of the solid-liquid cluster increases. There may be some experimental evidence for such a scenario: Koga et al \cite{Koga04} have recently reported the observation of an icosahedral to decahedral transition in free gold particles as they were annealed close to their melting temperatures. It was suggested that this was a thermally activated transition driven by cluster thermodynamics. However, it is quite possible that the gold particles were in a coexisting solid-liquid state at temperatures just below the melting point, and in this case the solid-liquid interface may well have played a key role in the structural transition. 

\begin{figure}
\resizebox{\columnwidth}{!}{\includegraphics{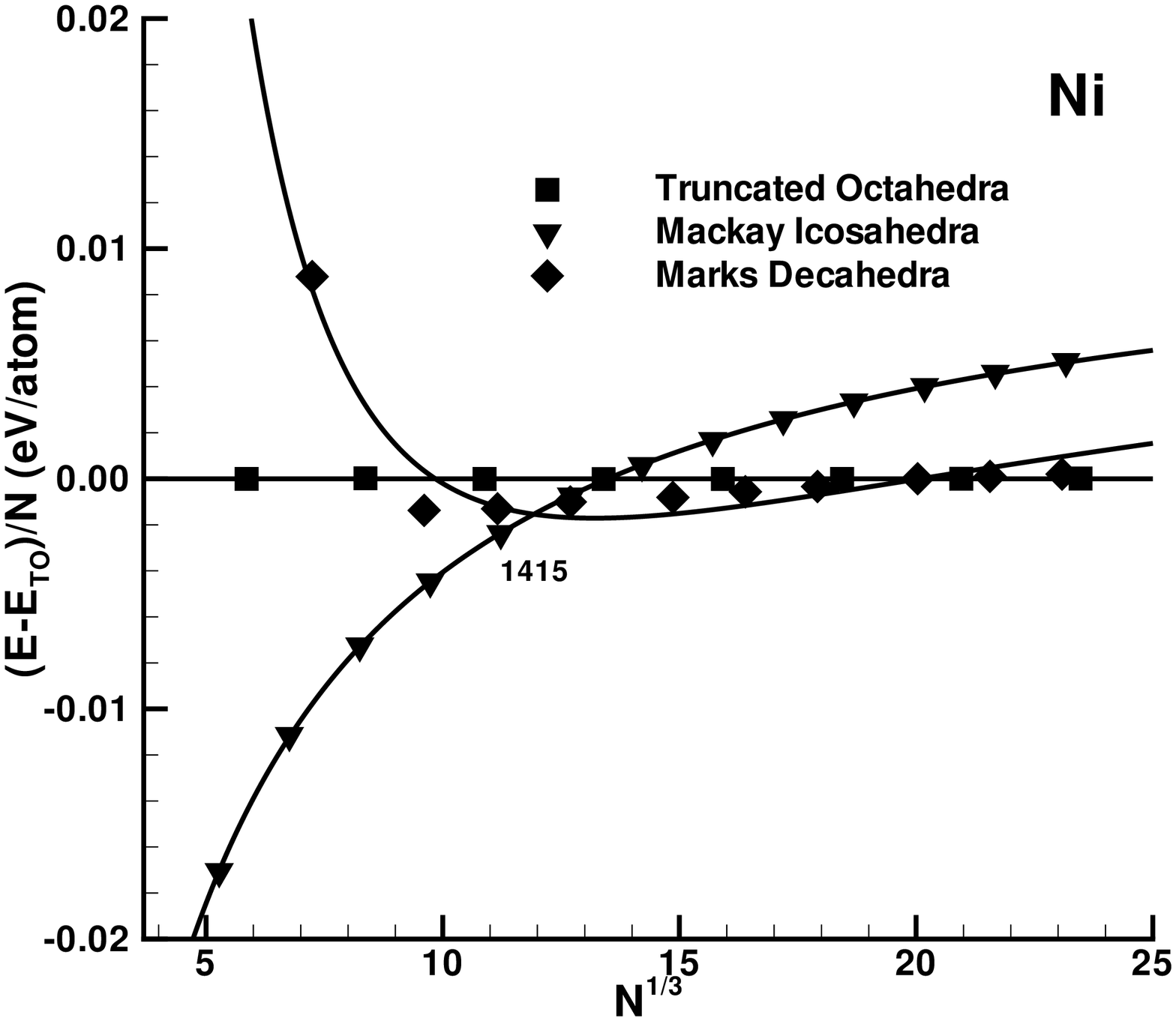}}
\caption{\label{fig1} Comparison of the relaxed zero temperature energies of icosahedra, Marks decahedra and truncated octahedra for nickel. Energies are given relative to a fit to the energies of the truncated octahedra sequence.}
\end{figure}

In this letter we report the transition of the solid region in a coexisting solid-liquid nickel cluster from a partial
icosahedron to a decahedron. We have applied molecular dynamics simulations to construct a caloric curve for a 1415-atom
cluster using an embedded atom method (EAM) potential \cite{Foiles86}. Initially we attempted to identify the stable structure at zero temperature by comparing the potential energies of closed-shell truncated octahedra (including variants such as the (TO)$^-$ and (TO)$^+$ structures \cite{Cleveland99}), cuboctahedra, icosahedra, and Marks \cite{Marks94} and Ino decahedra \cite{Ino66}. In figure~\ref{fig1} we show the relaxed energies for the icosahedra sequence and the Marks decahedra sequence relative to a fit to energies of the truncated octahedra sequence. In Ni we find that below 2869 atoms the icosahedral structures are lowest in energy, between 2869 and 12298 atoms the decahedral sequence is more stable and above 12298 atoms the fcc truncated octahedra become favorable. This is similar to the structural sequence predicted in Ref~\cite{Cleveland91} using the same potential, although there it was estimated that fcc structures would become stable at sizes of 17000 atoms. From figure~\ref{fig1}, we note that the icosahedron should be stable at a cluster size of 1415-atoms but that the 1389-atom Marks decahedron and the 1289-atom truncated octahedron are only slightly less favorable energetically. Thus a 1415-atom nickel cluster would seem to be a good candidate for observing structural transitions during solid-liquid coexistence.

A previous molecular dynamics study of solid-liquid coexistence in a 1289-atom nickel truncated octahedron \cite{Cleveland94} found partial wetting of the solid by the melt (with the solid exposing a solid lens consisting of (111)-facets) and then full wetting at temperatures closer to the melting point (with the solid completely covered by the melt). This study also used the same EAM potential as our simulations. Other molecular dynamics simulations of coexistence in metal clusters have studied copper \cite{Nielsen94}, gold \cite{Cleveland99} and lead \cite{Hendy05}.  

The caloric curves were constructed in the constant energy (microcanonical) ensemble using the following procedure: at each fixed total energy the cluster was equilibrated for 150000 time steps (where $\Delta t$ = 2 fs) and then the kinetic energy was averaged over a further 150000 steps to obtain a temperature. An energy increment of 0.6 meV/atom was used to adjust the total energy between simulations by a uniform scaling of the kinetic energy. This corresponds to a heating rate of 1 meV/atom/nanosecond. To identify and characterize solid-liquid coexistence, we follow Cleveland et al \cite{Cleveland94}, using the bimodality of the distribution of diffusion coefficients to distinguish solid and liquid atoms. We have previously used this method in Ref~\cite{Hendy05} to characterize the coexisting solid-liquid states in Pb clusters. 

\begin{figure}
\resizebox{\columnwidth}{!}{\includegraphics{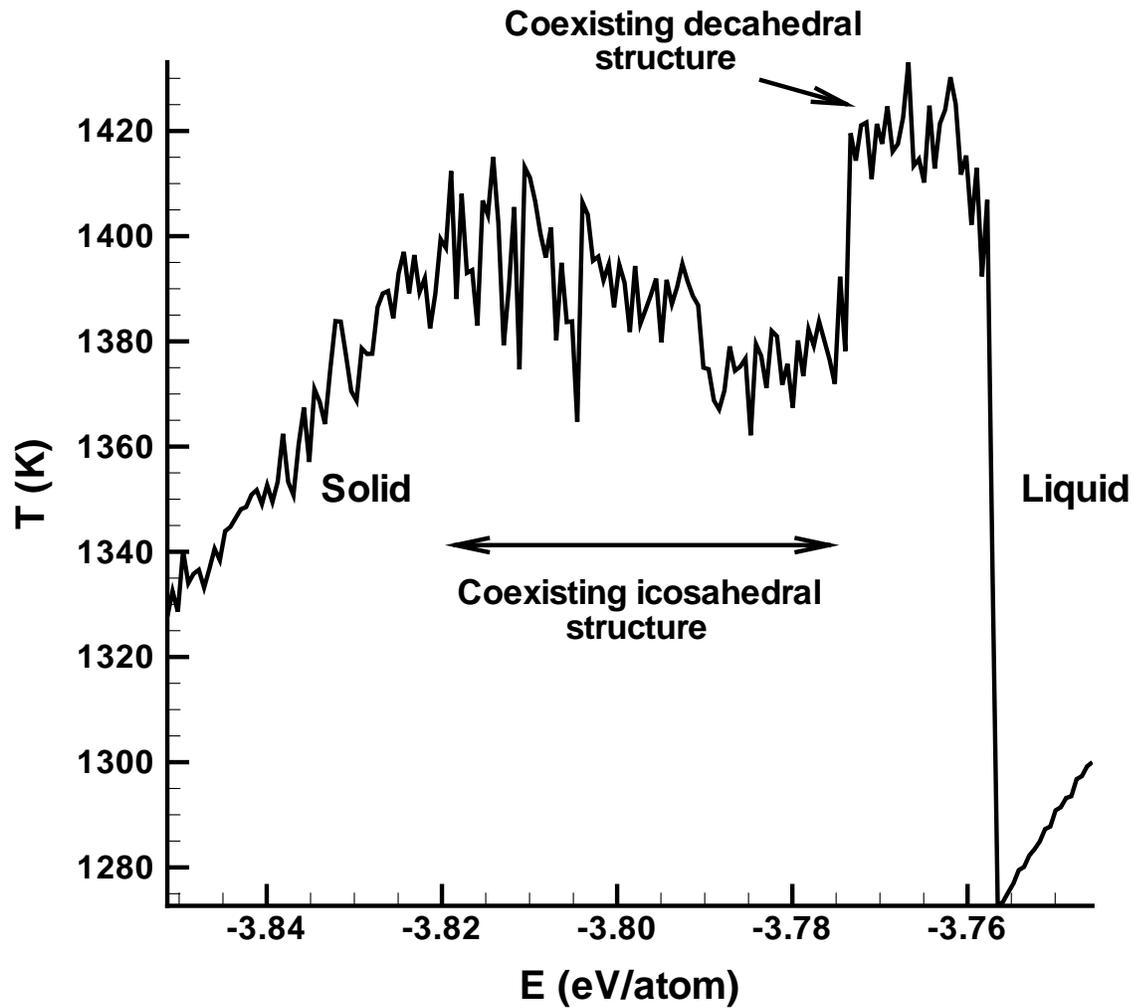}}
\caption{\label{fig2} Caloric curve for the 1415-atom icosahedron. The onset of coexistence occurs at an
energy of approximately $E=-3.81$ eV/atom. There is a second transition at $E=-3.77$ eV/atom, followed finally
by melting at $E=-3.755$ eV/atom.}
\end{figure}

\begin{figure}
\resizebox{\columnwidth}{!}{\includegraphics{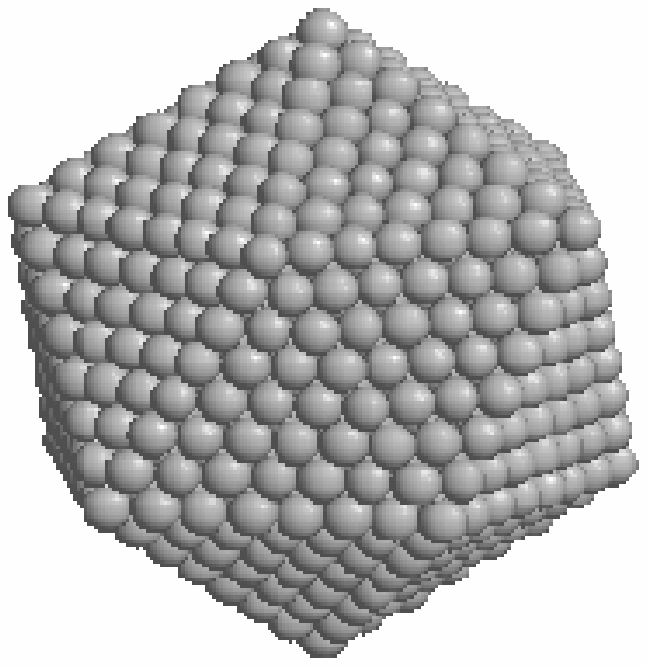} \includegraphics{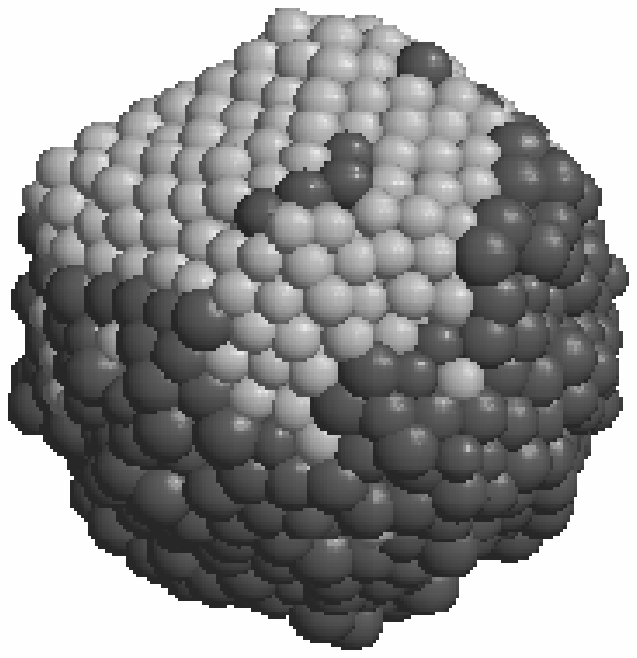} \includegraphics{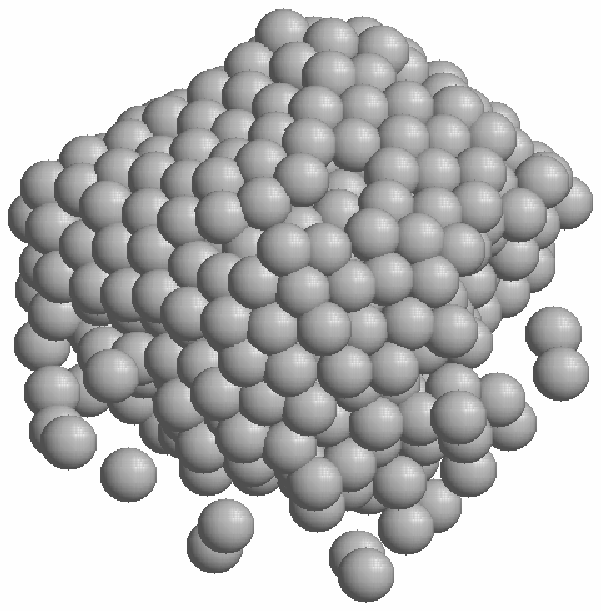}}
\caption{\label{fig3} The figures show icosahedron prior to heating (left) and the coexisting solid-liquid icosahedron  at $E=-3.78$ eV/atom just prior to the transition (center and right). In the center the liquid atoms are shown in a darker shade. Note that the liquid only partially wets the solid, exposing (111)-facets, whereas if the liquid completely wet the solid then there would be no solid exposed. On the right, the liquid atoms have been removed to show only the solid.}
\end{figure}

Figure~\ref{fig2} shows the resulting caloric curve for the 1415-atom icosahedron. At total energies below $E=-3.82$
eV/atom the cluster is fully solid. The onset of coexistence occurs at energies above just this. For energies between
$E=-3.82$ eV/atom and $E=-3.77$ eV/atom the structure of the cluster is that of an incomplete icosahedron partially wetted by a region of melt: figure~\ref{fig3} compares the icosahedron structure prior to heating with snapshots of the coexisting cluster at $E=-3.78$ eV/atom with the liquid atoms shaded in dark (center), and then removed (right). In the coexisting structure it is possible to distinguish 6 nearly complete tetrahedra of the 20 that make up a full icosahedron. This structure is very similar to the icosahedral solid-liquid structures seen in molecular dynamics simulations of lead clusters \cite{Hendy05}. We note that the region of exposed (111)-facets of the icosahedron do not resemble a lens in the sense of Ref~\cite{Cleveland94}, where it used to refer to a patch of exposed solid surrounded by liquid in a nearly wetted cluster. 

\begin{figure}
\resizebox{\columnwidth}{!}{\includegraphics{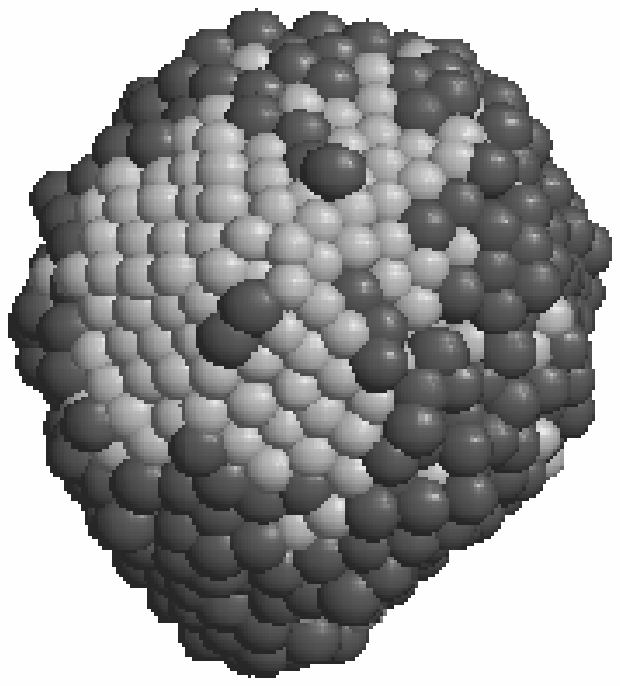} \includegraphics{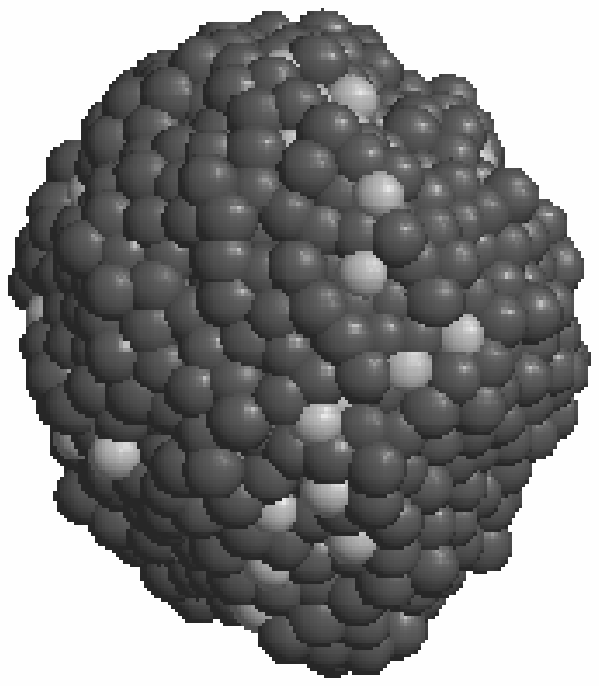}}
\resizebox{\columnwidth}{!}{\includegraphics{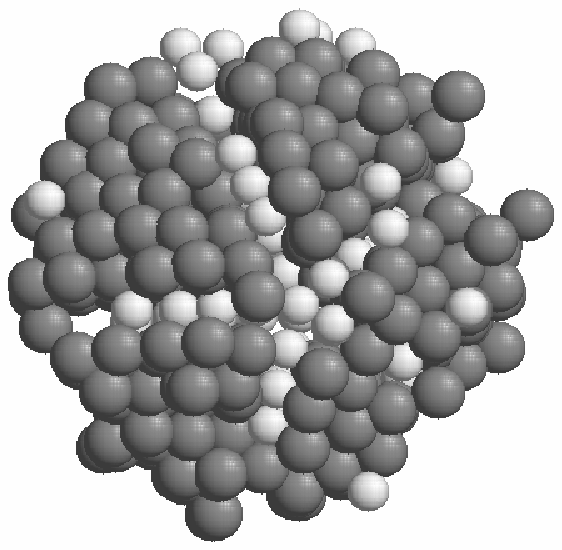} \includegraphics{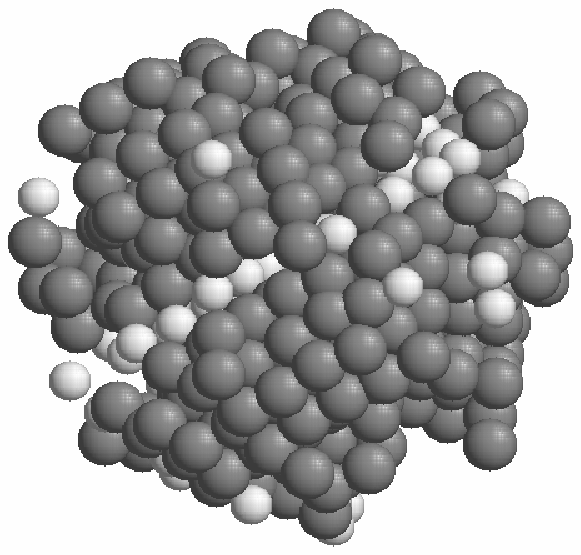}}
\caption{\label{fig4} The plots show the coexisting solid-liquid decahedron at $E=-3.77$ eV/atom just after the transition. In the top two snapshots, taken from opposite viewpoints, the liquid atoms are shown in a darker shade.  
In the bottom snapshots, the liquid atoms have been removed and CNA analysis has been used to highlight the twin planes of the decahedron. On the bottom left the viewpoint is down the fivefold axis of the decahedron. On the right the picture is looking side on to one of the twin planes.}
\end{figure}

At approximately $E=-3.77$ eV/atom a transition occurs. This is visible in the caloric curve by the sudden increase
in temperature at this energy indicating that the cluster has lowered its potential energy. Note that the snapshot in
figure~\ref{fig3} is taken just prior to this transition at $E=-3.78$ eV/atom. In figure~\ref{fig4} we show snapshots from immediately after the transition. The top left snapshot shows exposed (111)-facets of the solid; the top right snapshot, looking at the opposite side of the cluster, shows no exposed solid. The region of exposed crystal facets is much more lens-like than in the case of the icosahedron with the melt appear to wet the solid more completely, although analysis of the distribution of diffusion coefficients suggests the liquid fraction has dropped slightly. In the bottom pictures, a common neighbor analysis \cite{CNA93} (CNA) has been performed using the index classification from Ref~\cite{Hendy02} to identify the structure of the solid region. The liquid atoms have been removed to show only atoms with fcc or hcp symmetries. Two angles are shown: looking down the fivefold axis of the solid (bottom left) and looking side on to this axis down one of the twin planes (bottom right). The structure closely resembles that of a five-shell Marks decahedron \cite{Marks94}. We conclude the that transition seen at $E=-3.77$ eV/atom in the caloric curve (figure~\ref{fig2}) is a transition of the solid from an incomplete icosahedron to a complete decahedron.

\begin{figure}
\resizebox{\columnwidth}{!}{\includegraphics{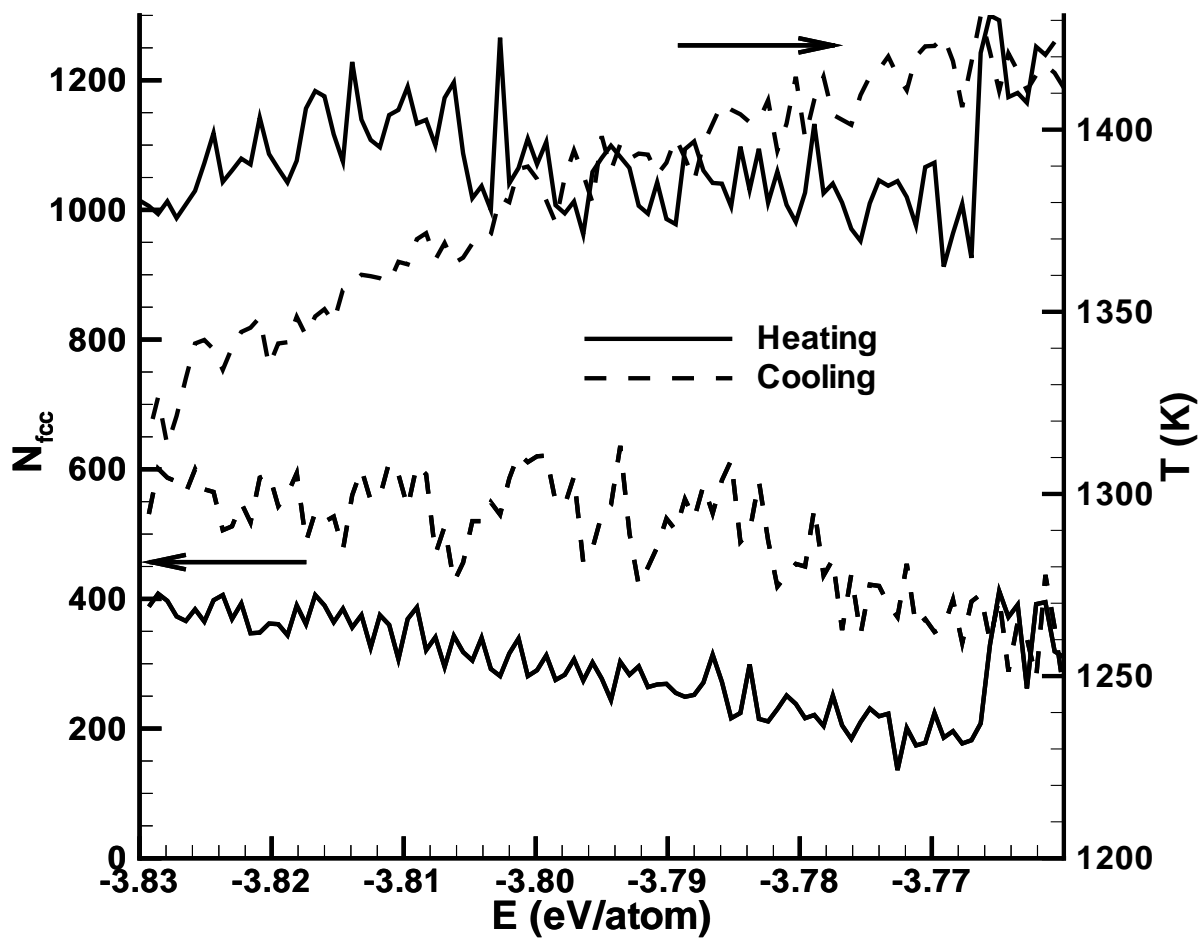}}
\caption{\label{fig5} The plot shows a second caloric curve where the cluster is heated (solid lines) until the icosahedral to decahedral transition and then cooled (dashed lines). Also shown is the corresponding the number of fcc atoms (as calculated by CNA analysis) 
in the cluster.}
\end{figure}

Figure~\ref{fig5} shows a second caloric curve constructed by first heating from $E=-3.83$ eV/atom to $E=-3.76$ eV/atom
and then cooling back to $E=-3.83$ eV/atom at the same rate. The transition from icosahedron to decahedron occurs at $E=-3.766$ eV/atom, which is close to the transition in the first caloric curve (figure~\ref{fig1}) suggesting that this is a good estimate of the transition energy. However we note in figure~\ref{fig5} that upon cooling there is no transition back to the icosahedron. Presumably, the decahedron is kinetically trapped, a common occurrence in both simulations and real clusters \cite{Koga04}. The the decahedron is fully solid at approximately $E=-3.79$ eV/atom. Note that the temperature of the decahedron falls below that of the icosahedron at $E=-3.80$ eV/atom; at this energy both clusters are solid indicating that the solid icosahedral structure has a lower potential energy than the decahedron consistent with the zero temperature calculations (figure~\ref{fig1}). 

Also shown in figure~\ref{fig5} are the number of atoms in bulk fcc positions (obtained via CNA analysis of the cluster structure) as the cluster is heated and cooled. During the heating phase the number of bulk fcc atoms steadily declines (as the solid fraction of the icosahedron decreases) until the transition at $E=-3.766$ eV/atom. At this point the number of bulk fcc atoms ($N_{fcc}$) jumps sharply as the transition to the decahedral structure occurs. Upon cooling $N_{fcc}$ increases (as the solid fraction increases) until coexistence ceases at about $E=-3.79$ eV/atom where it can be seen that $N_{fcc}$ becomes relatively static. In a perfect 1389-atom Marks decahedron (that is a relaxed structure at zero temperature) CNA analysis counts $N_{fcc}=650$ and in a perfect 1415-atom icosahedron CNA analysis counts $N_{fcc}=400$. Both the decahedral and icosahedral structure approach these values by $E=-3.83$ eV/atom.

\begin{figure}
\resizebox{\columnwidth}{!}{\includegraphics{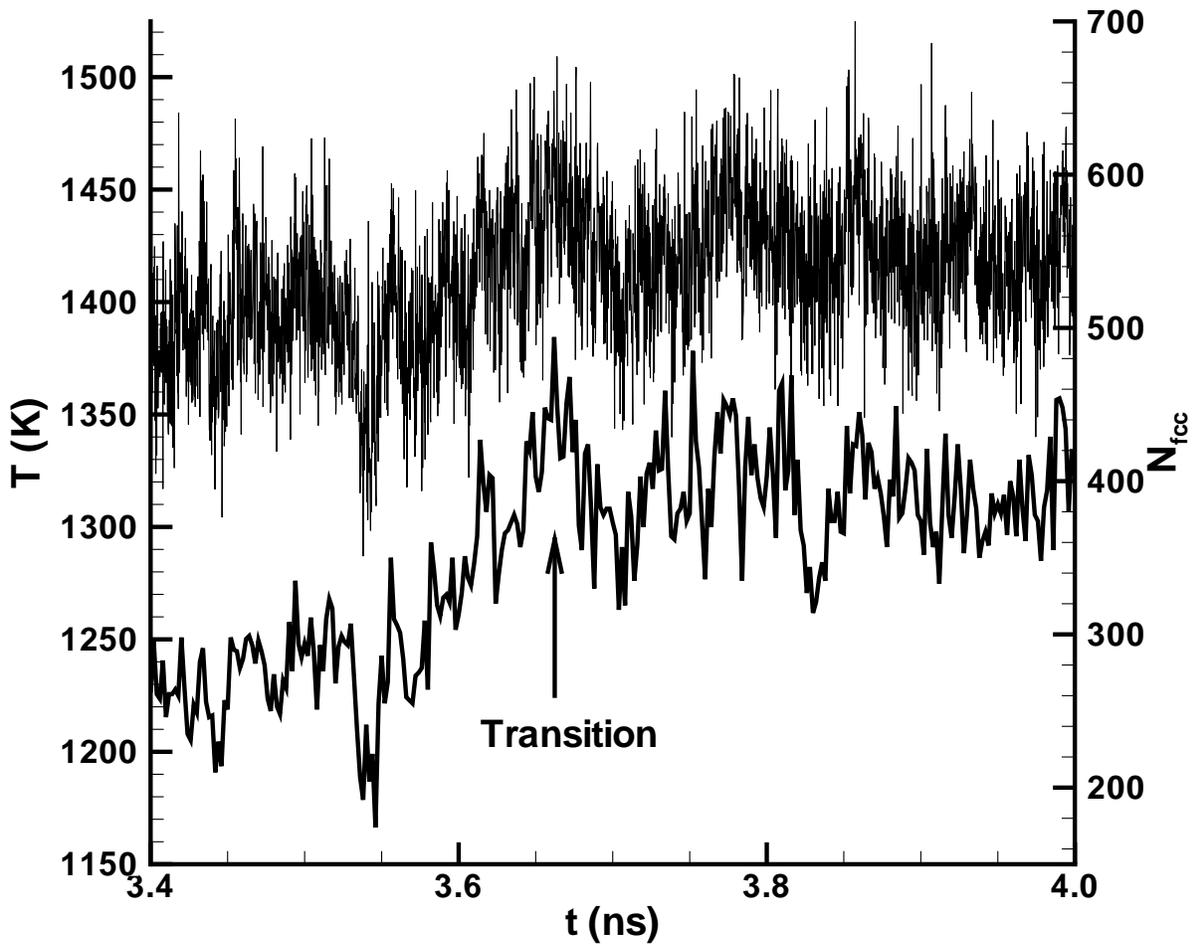}}
\caption{\label{fig6} The plot shows the time evolution of the temperature and number of fcc atoms in the cluster as
it goes through the transition from icosahedron to decahedron at $E=-3.77$ (eV/atom).}
\end{figure}

To test whether the location of the transition depends on cooling rate we conducted longer 4 ns (2 million steps) constant energy simulations of the icosahedron at several energies near the transition point in  the caloric curve. In figure~\ref{fig6} we show the time evolution of the temperature and the number of fcc atoms in the $E=-3.77$ eV/atom simulation. Here the transition is seen to occur at $t=3.65$ ns where an increase in $N_{fcc}$ and a jump in the temperature is visible. We see no such transition in 4 ns simulations at $E=-3.78$ eV/atom and $E=-3.775$ eV/atom. This again suggests that we have a good estimate of the transition energy.   

There are several known instances where solid clusters undergo structural transitions prior to melting. Small gold clusters have been seen in simulations to undergo transitions from truncated-octahedral structures, which are globally stable at zero temperature, to icosahedral structures at energies just below the melting point \cite{Cleveland98}. Similarly a study of icosahedral Morse clusters found complex surface reconstructions prior to melting \cite{Doye97}. In these cases, the transitions are driven by the thermodynamics of the solid phase rather than contact with a melt. However it is unlikely that similar thermodynamics are driving the transition seen here as the transition is accompanied by an decrease in potential energy of about 10 meV/atom (corresponding to the rise in temperature that can be seen in figure~\ref{fig2}). This decrease in potential energy of the overall cluster comes from the contribution of surface atoms which drop on average by 40 meV/atom after the transition; it seems likely that this comes from an improvement in the solid-liquid interfacial energy associated with the wetting of the decahedral structure. We note that the interior atoms actually experience a net increase in potential energy of approximately 10 meV/atom. As remarked earlier, a previous study of phase coexistence in Ni clusters found a preference for fcc cuboctahedral cluster to expose (111)-facets (lenses) to the vacuum or vapour \cite{Cleveland94}, suggesting a preference for the melt to wet the (100)-facets. Molten lead is known to have a preference to wet crystalline lead (100)-facets over the (111)-facets \cite{Pluis87} and the structure of the coexisting icosahedron in figure~\ref{fig3} strongly resembles the coexisting icosahedron structure seen in simulations of lead \cite{Hendy05}. It is possible that the higher-energy (100)-facets on the decahedron offer an improved solid-liquid interfacial energy over the (111)-facets; this is hard to verify directly
due to the dificulty in defining the solid-liquid interface in such a small system.  
  
At cluster sizes near 1400 atoms, the EAM potential \cite{Foiles86} predicts that several structures possess very similar energies, including the icosahedron and the decahedron (with a size of 1389 atoms). Indeed the transition occurs between an incomplete icosahedron and a decahedron. In simulations of Ni clusters at other sizes \cite{Hendy05a}, we have observed static solid-liquid coexistence in a 923-atom icosahedron (but not in cluster sizes below this) and we did not observe this transition to a decahedral structure.   

We conclude that we have found a structural transition that occurs in the solid part of a coexisting solid-liquid 
nickel cluster modelled using an EAM potential. We believe this is the first suggestion that structural changes can occur in solid clusters to accommodate a coexisting melt. This effect may explain the recent observation of a icosahedral to decahedral transition seen in free gold particles \cite{Koga04} as they were annealed close to the  
melting point and it could provide an important kinetic mechanism for controlling the structure of nanoscale metal nanoparticles.

\begin{acknowledgments}
The authors would like to acknowledge financial support from the MacDiarmid Institute for Nanotechnology and Advanced Materials.
\end{acknowledgments}


\clearpage
\thispagestyle{empty}

\clearpage
\thispagestyle{empty}

\clearpage
\thispagestyle{empty}

\clearpage
\thispagestyle{empty}

\clearpage
\thispagestyle{empty}

\clearpage
\thispagestyle{empty}

\begin{thebibliography}{27}
\expandafter\ifx\csname natexlab\endcsname\relax\def\natexlab#1{#1}\fi
\expandafter\ifx\csname bibnamefont\endcsname\relax
  \def\bibnamefont#1{#1}\fi
\expandafter\ifx\csname bibfnamefont\endcsname\relax
  \def\bibfnamefont#1{#1}\fi
\expandafter\ifx\csname citenamefont\endcsname\relax
  \def\citenamefont#1{#1}\fi
\expandafter\ifx\csname url\endcsname\relax
  \def\url#1{\texttt{#1}}\fi
\expandafter\ifx\csname urlprefix\endcsname\relax\def\urlprefix{URL }\fi
\providecommand{\bibinfo}[2]{#2}
\providecommand{\eprint}[2][]{\url{#2}}

\bibitem[{\citenamefont{Ino}(1966)}]{Ino66}
\bibinfo{author}{\bibfnamefont{S.}~\bibnamefont{Ino}}, \bibinfo{journal}{J.
  Phys. Soc. Jpn} \textbf{\bibinfo{volume}{27}}, \bibinfo{pages}{941}
  (\bibinfo{year}{1967}).
  
\bibitem[{\citenamefont{Pochon et al}(2004)}]{Pochon04}
  \bibinfo{author}{\bibfnamefont{S.} \bibnamefont{Pochon}},
  \bibinfo{author}{\bibfnamefont{K.~F.} \bibnamefont{MacDonald}},
  \bibinfo{author}{\bibfnamefont{R.~J.} \bibnamefont{Knize}} \bibnamefont{and}
  \bibinfo{author}{\bibfnamefont{N.~I.} \bibnamefont{Zheludev}},
  \bibinfo{journal}{Phys. Rev. Lett.} \textbf{\bibinfo{volume}{92}},
  \bibinfo{pages}{145702} (\bibinfo{year}{2004}).

\bibitem[{\citenamefont{Lee and Mori}(2004)}]{Lee04}
  \bibinfo{author}{\bibfnamefont{J.-G.} \bibnamefont{Lee}} \bibnamefont{and}
  \bibinfo{author}{\bibfnamefont{H.} \bibnamefont{Mori}},
  \bibinfo{journal}{Phys. Rev. B.} \textbf{\bibinfo{volume}{70}},
  \bibinfo{pages}{144105} (\bibinfo{year}{2004}).
  
\bibitem[{\citenamefont{Reiss et~al.}(1988)}]{Reiss88}
\bibinfo{author}{\bibfnamefont{H.} \bibnamefont{Reiss}}
\bibinfo{author}{\bibfnamefont{P.} \bibnamefont{Mirabel}},
\bibnamefont{and}
\bibinfo{author}{\bibfnamefont{R.~L.} \bibnamefont{Whetten}},
\bibinfo{journal}{J. Phys. Chem.} \textbf{\bibinfo{volume}{92}},
  \bibinfo{pages}{7241-7246} (\bibinfo{year}{1988}).

\bibitem[{\citenamefont{Wales and Berry}(1994)}]{Wales94}
  \bibinfo{author}{\bibfnamefont{D.~J.} \bibnamefont{Wales}} \bibnamefont{and}
  \bibinfo{author}{\bibfnamefont{R.~S.} \bibnamefont{Berry}},
  \bibinfo{journal}{Phys. Rev. Lett.} \textbf{\bibinfo{volume}{73}},
  \bibinfo{pages}{2875} (\bibinfo{year}{1994}).
  
  \bibitem[{\citenamefont{Honeycutt and Andersen}(1987)}]{Honeycutt87}
\bibinfo{author}{\bibfnamefont{J.~D.} \bibnamefont{Honeycutt}}
\bibnamefont{and}
\bibinfo{author}{\bibfnamefont{H.~C.} \bibnamefont{Andersen}},
\bibinfo{journal}{J. Phys. Chem.} \textbf{\bibinfo{volume}{91}},
  \bibinfo{pages}{4950} (\bibinfo{year}{1987}).

\bibitem[{\citenamefont{Lynden-Bell and Wales}(1994)}]{Lynden-Bell}
  \bibinfo{author}{\bibfnamefont{R.~M.} \bibnamefont{Lynden-Bell}}
  \bibnamefont{and} \bibinfo{author}{\bibfnamefont{D.~J.} \bibnamefont{Wales}},
  \bibinfo{journal}{J. Chem. Phys.} \textbf{\bibinfo{volume}{101}},
  \bibinfo{pages}{1460} (\bibinfo{year}{1994}).
  
\bibitem[{\citenamefont{Schmidt et~al.}(2001)}]{Schmidt01}
\bibinfo{author}{\bibfnamefont{M.} \bibnamefont{Schmidt}},
\bibinfo{author}{\bibfnamefont{R.} \bibnamefont{Kusche}},
\bibinfo{author}{\bibfnamefont{T.} \bibnamefont{Hippler}},
\bibinfo{author}{\bibfnamefont{J.} \bibnamefont{Donges}},
\bibinfo{author}{\bibfnamefont{W.} \bibnamefont{Kronmuller}},
\bibinfo{author}{\bibfnamefont{B.} \bibnamefont{von Issendorff}},
\bibnamefont{and}
\bibinfo{author}{\bibfnamefont{H.} \bibnamefont{Haberland}},
\bibinfo{journal}{Phys. Rev. Lett.} \textbf{\bibinfo{volume}{86}},
  \bibinfo{pages}{1191-1194} (\bibinfo{year}{2001}).

\bibitem[{\citenamefont{Hendy}(2005)}]{Hendy05}
  \bibinfo{author}{\bibfnamefont{S.~C.} \bibnamefont{Hendy}},
  \bibinfo{journal}{Phys. Rev. B} \textbf{\bibinfo{volume}{71}},
  \bibinfo{pages}{115404} (\bibinfo{year}{2005}).

\bibitem[{\citenamefont{Breaux et al}(2005)}]{Breaux05}
  \bibinfo{author}{\bibfnamefont{G.~A.} \bibnamefont{Breaux}},
  \bibinfo{author}{\bibfnamefont{C.~M.} \bibnamefont{Neal}},
  \bibinfo{author}{\bibfnamefont{B.} \bibnamefont{Cao}}
  \bibnamefont{and}
  \bibinfo{author}{\bibfnamefont{M.~F.} \bibnamefont{Jarrold}},
  \bibinfo{journal}{Phys. Rev. Lett.} \textbf{\bibinfo{volume}{94}},
  \bibinfo{pages}{173401} (\bibinfo{year}{2005}).
  
\bibitem[{\citenamefont{Foiles et al}(1986)}]{Foiles86}
\bibinfo{author}{\bibfnamefont{S.~M.} \bibnamefont{Foiles}},
\bibinfo{author}{\bibfnamefont{M.~I.}~\bibnamefont{Baskes}}
\bibnamefont{and}
\bibinfo{author}{\bibfnamefont{M.~S.}~\bibnamefont{Daw}},
\bibinfo{journal}{Phys. Rev. B} \textbf{\bibinfo{volume}{33}},
\bibinfo{pages}{7983-7991} (\bibinfo{year}{1986}).

\bibitem[{\citenamefont{Cleveland et~al.}(1999)\citenamefont{Cleveland,
  Luedtke and Landman}}]{Cleveland99}
\bibinfo{author}{\bibfnamefont{C.~L.} \bibnamefont{Cleveland}},
\bibinfo{author}{\bibfnamefont{W.~D.}~\bibnamefont{Luedtke}}
\bibnamefont{and}
\bibinfo{author}{\bibfnamefont{U.}~\bibnamefont{Landman}},
\bibinfo{journal}{Phys. Rev. B} \textbf{\bibinfo{volume}{60}},
\bibinfo{pages}{5065-5077} (\bibinfo{year}{1999}).

\bibitem[{\citenamefont{Marks}(2002)}]{Marks94}
 \bibinfo{author}{\bibfnamefont{L.~D.} \bibnamefont{Marks}},
  \bibinfo{journal}{Rep. Prog. Phys.}  \textbf{\bibinfo{volume}{57}},
  \bibinfo{pages}{603} (\bibinfo{year}{1994}).  
  
\bibitem[{\citenamefont{Cleveland and Landman}(1991)\citenamefont{Cleveland and Landman}}]{Cleveland91}
\bibinfo{author}{\bibfnamefont{C.~L.} \bibnamefont{Cleveland}}
\bibnamefont{and}
\bibinfo{author}{\bibfnamefont{U.}~\bibnamefont{Landman}},
\bibinfo{journal}{J. Chem. Phys.} \textbf{\bibinfo{volume}{94}},
\bibinfo{pages}{7376-7396} (\bibinfo{year}{1991}).

\bibitem[{\citenamefont{Cleveland et~al.}(1994)\citenamefont{Cleveland,
  Luedtke and Landman}}]{Cleveland94}
\bibinfo{author}{\bibfnamefont{C.~L.} \bibnamefont{Cleveland}},
\bibinfo{author}{\bibfnamefont{U.}~\bibnamefont{Landman}} \bibnamefont{and}
\bibinfo{author}{\bibfnamefont{W.~D.}~\bibnamefont{Luedtke}},
\bibinfo{journal}{J. Phys. Chem.} \textbf{\bibinfo{volume}{98}},
\bibinfo{pages}{6272-6279} (\bibinfo{year}{1994}).

\bibitem[{\citenamefont{Nielsen et~al.}(1994)}]{Nielsen94}
\bibinfo{author}{\bibfnamefont{O.~H.} \bibnamefont{Nielsen}},
\bibinfo{author}{\bibfnamefont{J.~P.} \bibnamefont{Sethna}},
\bibinfo{author}{\bibfnamefont{P.} \bibnamefont{Stoltze}},
\bibinfo{author}{\bibfnamefont{K.~W.} \bibnamefont{Jacobsen}},
\bibnamefont{and}
  \bibinfo{author}{\bibfnamefont{J.~K.} \bibnamefont{Norskov}},
  \bibinfo{journal}{Europhys. Lett.} \textbf{\bibinfo{volume}{26}},
  \bibinfo{pages}{51-56} (\bibinfo{year}{1994}).

\bibitem[{\citenamefont{Clarke and Jonsson}(1993)}]{CNA93}
\bibinfo{author}{\bibfnamefont{A.~S.} \bibnamefont{Clarke}} \bibnamefont{and}
  \bibinfo{author}{\bibfnamefont{H.}~\bibnamefont{Jonsson}},
  \bibinfo{journal}{Phys. Rev. E} \textbf{\bibinfo{volume}{47}},
  \bibinfo{pages}{3975} (\bibinfo{year}{1993}).
  
\bibitem[{\citenamefont{Hendy and Doye}(2002)}]{Hendy02}
 \bibinfo{author}{\bibfnamefont{S.~C.} \bibnamefont{Hendy}} \bibnamefont{and}
\bibinfo{author}{\bibfnamefont{J.~P.~K.} \bibnamefont{Doye}}  ,
  \bibinfo{journal}{Phys. Rev. B}  \textbf{\bibinfo{volume}{66}},
  \bibinfo{pages}{235402} (\bibinfo{year}{2002}).

\bibitem[{\citenamefont{Koga et~al}(2004)}]{Koga04}
\bibinfo{author}{\bibfnamefont{K.} \bibnamefont{Koga}},
\bibinfo{author}{\bibfnamefont{T.} \bibnamefont{Ikeshoji}} 
  \bibnamefont{and}
\bibinfo{author}{\bibfnamefont{K.~I.} \bibnamefont{Sugawara}}  ,
  \bibinfo{journal}{Phys. Rev. Lett.}  \textbf{\bibinfo{volume}{92}},
  \bibinfo{pages}{115507} (\bibinfo{year}{2004}).
  
\bibitem[{\citenamefont{Cleveland et~al}(1998)\citenamefont{Cleveland,
  Luedtke and Landman}}]{Cleveland98}
\bibinfo{author}{\bibfnamefont{C.~L.} \bibnamefont{Cleveland}},
\bibinfo{author}{\bibfnamefont{W.~D.}~\bibnamefont{Luedtke}},
\bibnamefont{and}
\bibinfo{author}{\bibfnamefont{U.}~\bibnamefont{Landman}},
\bibinfo{journal}{Phys. Rev. Lett.} \textbf{\bibinfo{volume}{81}},
\bibinfo{pages}{2036-2039} (\bibinfo{year}{1998}).

\bibitem[{\citenamefont{Doye and Wales}(1997)}]{Doye97}
\bibinfo{author}{\bibfnamefont{J.~P.~K.} \bibnamefont{Doye}} \bibnamefont{and}
  \bibinfo{author}{\bibfnamefont{D.~J.} \bibnamefont{Wales}},
  \bibinfo{journal}{Zeit. Phys. D} \textbf{\bibinfo{volume}{40}},
  \bibinfo{pages}{466} (\bibinfo{year}{1997}).

\bibitem[{\citenamefont{Pluis et al}(1987)}]{Pluis87}
\bibinfo{author}{\bibfnamefont{B.} \bibnamefont{Pluis}},
\bibinfo{author}{\bibfnamefont{A.~W.}~\bibnamefont{Denier van der Gon}},
\bibinfo{author}{\bibfnamefont{J.~W.~M.}~\bibnamefont{Frenken}}
\bibnamefont{and}
\bibinfo{author}{\bibfnamefont{J.~F.}~\bibnamefont{van der Veen}},
\bibinfo{journal}{Phys. Rev. Lett.} \textbf{\bibinfo{volume}{59}},
\bibinfo{pages}{2678-2681} (\bibinfo{year}{1987}).
  
\bibitem[{\citenamefont{Schebarchov and Hendy}(2005)}]{Hendy05a}
\bibinfo{author}{\bibfnamefont{D.}~\bibnamefont{Schebarchov}} \bibnamefont{and}
\bibinfo{author}{\bibfnamefont{S.~C.}~\bibnamefont{Hendy}},
  to appear in \bibinfo{journal}{J. Chem. Phys.} (\bibinfo{year}{2005}).
\end{thebibliography}
\end{document}